\documentclass[11pt]{amsart}
\usepackage{epsfig}
\usepackage{amsbsy,latexsym}
\usepackage{amsmath}
\usepackage{amssymb, mathrsfs}
\usepackage[mathscr]{eucal}
\usepackage{hyperref}
\usepackage{amsfonts}
\usepackage{amsthm}
\usepackage{amsmath}
\usepackage{amsfonts}
\usepackage{latexsym}
\usepackage{amssymb}
\usepackage{amscd}
\usepackage[latin1]{inputenc}
\usepackage{verbatim}
\usepackage[english]{babel}

\newtheorem{definition}{Definition}[section]
\newtheorem{theorem}[definition]{Theorem}

\newtheorem{corollary}[definition]{Corollary}
\newtheorem{proposition}[definition]{Proposition}
\theoremstyle{definition}
\newtheorem{remark}[definition]{Remark}

\newcommand\bk{\style B(\style K)}     %%%% B(K)

\newcommand\tr{ \operatorname{Tr} }

\def\ket#1{| #1 \rangle}
\def\bra#1{\langle #1 |}
\def\kb#1#2{|#1\rangle\!\langle #2 |}
\def\bk#1#2{\langle #1 |#2\rangle}

\title[Entanglement Breaking Channels, Stochastic Matrices, Primitivity]{Entanglement Breaking Channels, Stochastic Matrices, and Primitivity}

\begin{document}

\author[J.~Ahiable, D.~W. Kribs, J.~Levick, R. Pereira, M.~Rahaman]{Jennifer~Ahiable$^{1}$, David~W.~Kribs$^{1,2}$, Jeremy Levick$^{1,2}$, Rajesh Pereira$^{1}$, Mizanur Rahaman$^{3}$}

\address{$^1$Department of Mathematics \& Statistics, University of Guelph, Guelph, ON, Canada N1G 2W1}
\address{$^2$Institute for Quantum Computing, University of Waterloo, Waterloo, ON, Canada N2L 3G1}
%\address{$^2$Department of Mathematics \& Applied Mathematics, University of Cape Town, Cape Town 7700, South Africa}
%\address{$^4$Department of Pure Mathematics, University of Waterloo, Waterloo, ON, Canada N2L 3G1}
\address{$^3$Department of Mathematics, BITS Pilani K. K Birla Goa Campus,  Goa 403726 India}



%%%%%%%%%%%%%%%%%%% title and abstract %%%%%%%%%%%%%%%%%%%%%%%%%%%%%%

\begin{abstract}
We consider the important class of quantum operations (completely positive trace-preserving maps) called entanglement breaking channels. We show how every such channel induces stochastic matrix representations that have the same non-zero spectrum as the channel. We then use this to investigate when entanglement breaking channels are primitive, and prove this depends on primitivity of the matrix representations. This in turn leads to tight bounds on the primitivity index of entanglement breaking channels in terms of the primitivity index of the associated stochastic matrices. We also present examples and discuss open problems generated by the work.
\end{abstract}
%%%%%%%%%%%%%%%%%%%%%%%

\maketitle

\section{Introduction}

Concepts and tools from linear algebra and matrix theory have played a role in the field of quantum information theory since its inception years ago. Over time, this role has grown along with the intersection between these fields \cite{holevo2012quantum,nielsen}. One area of fundamental importance in this respect is quantum entanglement theory, one of the most challenging subjects within quantum information and more generally in modern science. The study of quantum entanglement has benefited from the application and development of matrix theoretic techniques from its beginnings, with many examples including some recent works involving the authors given by \cite{girard2020mixed,kribs2021nullspaces,kribs2020vector,LKP,pandey2020entanglement}.

In this paper, we contribute to this study with an investigation of an important class of quantum operations, which are given mathematically by completely positive trace-preserving maps on matrices, called entanglement breaking channels \cite{holevo1998coding,horodecki2003entanglement}. We bring two key notions from matrix theory to their study by identifying stochastic matrix representations for the channels, and, built on this, we conduct an analysis of the primitivity \cite{rahaman2019new,sanz2010quantum} of such channels based on corresponding matrix primitivity \cite{horn2012matrix,meyer2000matrix}. More specifically, we show how every so-called Holevo form of an entanglement breaking channel induces a certain stochastic matrix representation that has the same non-zero spectrum as the channel. We then prove that primitivity of the channel depends on primitivity of its matrix representations, and we use this to obtain tight bounds on the primitivity index of the channel in terms of the primitivity index of the associated stochastic matrices. We also present a number of examples, and finish with a discussion on some possible new directions to pursue coming out of this work.

This paper is organized as follows. In the next section we present requisite preliminary notions. The third section introduces the stochastic matrix representations of the channels. The fourth section contains the primitivity analysis. The final section includes our concluding remarks.

\section{Preliminaries}

We denote the set of complex $n\times n$ matrices, for a fixed $n\geq 1$, by $M_n (\mathbb{C})$. We will use the Dirac bra-ket notation for vectors, which labels a given fixed orthonormal basis for $\mathbb{C}^n$ as $\{ \ket{i} : 0 \leq i \leq n-1 \}$; the corresponding dual vectors as $\bra{i}$; and the outer product rank-one operators $E_{ij} := \kb{i}{j}$, which in matrix form is the matrix with a 1 in the $i,j$ entry and 0's elsewhere. The full set of $E_{ij}$ form a set of `matrix units' for the operator space $M_n (\mathbb{C})$ determined by the vectors $\{ \ket{i}\}$. The standard inner product $\bk{\phi}{\psi}$ on $\mathbb{C}^n$ we take as conjugate linear in the second argument (hence using the mathematics instead of the physics convention). The Hilbert-Schmidt trace inner product on $M_n (\mathbb{C})$ is given by $\bk{A}{B} = \tr(B^* A)$.

By a nonnegative $n\times n$ matrix, we mean a matrix whose entries are all (real) nonnegative numbers. We shall use the term stochastic matrix for non-negative matrices such that each column's entries from the matrix sum to 1; that is, each column forms a (classical) probability vector. (So we are using the term stochastic matrix for what are formally column stochastic matrices.)

A matrix $A\in M_n(\mathbb{C})$ is positive semidefinite if $\bk{A\psi}{\psi} \geq 0$ for all $\ket{\psi}\in \mathbb{C}^n$, and it is positive definite if this inequality is strict for all non-zero $\ket{\psi}$. Additionally will use the notation $\rho$ for density operators or density matrices, the quantum analogue of classical probability distributions; that is, positive semidefinite operators with trace equal to one.

Given a linear map $\Phi : M_n(\mathbb{C}) \rightarrow M_n(\mathbb{C})$, we will make use of its matrix representations in fixed bases for $\mathbb{C}^n$. Specifically, given an orthonormal basis $\{\ket{i} : 0 \leq i \leq n-1 \}$, we will consider the $n^2 \times n^2$ matrix $[\Phi]$ whose $(i_1,j_1)$, $(i_2,j_2)$ entry is $\bk{\Phi(E_{i_2j_2})}{E_{i_1j_1}} = \tr(E_{j_1i_1}\Phi(E_{i_2j_2}))$.

Quantum channels are central objects of study in quantum information \cite{holevo2012quantum,nielsen}, and are given mathematically by completely positive and trace preserving maps on $M_n (\mathbb{C})$. Every channel $\Phi$ can be represented in the `operator-sum form' by a set of operators $V_k$ called {\it Kraus operators}, represented as $n \times n$ matrices when an orthonormal basis for $\mathbb{C}^n$ is identified, such that $\Phi(\rho) = \sum_k V_k \rho V_k^*$ while satisfying the trace-preservation condition $\sum_k V^*_k V_k = I$. Here we have written $I$ for the $n$-dimensional identity operator, also viewing it when necessary as the $n\times n$ identity matrix. The {\it Choi matrix} \cite{choi1975cp} for a channel $\Phi$ will also make an appearance in our analysis, which is the matrix inside  $M_n (\mathbb{C}) \otimes M_n (\mathbb{C})$ given by $J(\Phi) = \sum_{i,j=1}^n E_{i,j} \otimes \Phi(E_{i,j})$. Many general properties of a channel can be determined from computable properties of its Choi matrix.

Quantum entanglement is a fundamental notion in quantum theory and we point the reader to other references such as \cite{holevo2012quantum,nielsen} for extensive introductions. An important class of channels are those that break all entanglement when acting on a composite system with the identity channel of the same size, $\mathrm{id}_n \otimes \Phi$  \cite{holevo1998coding,horodecki2003entanglement}. There are numerous equivalent characterizations of so-called {\it entanglement breaking channels}, including a physically motivated description as the composition of so-called quantum-classical and classical-quantum channels in the same orthonormal basis. They are also precisely the channels with an operator-sum representation comprised of rank one Kraus operators $V_k$, and the channels with separable Choi matrices. The description we shall focus on here, however, is the {\it Holevo form}:
\begin{equation}\label{holevo}
\Phi(\rho) = \sum_{k=1}^r \tr(F_k\rho) R_k,
\end{equation}
with each $R_k$ a density operator, and with the $\{F_k\}_{k=1}^r$ forming a positive-operator valued measure (POVM), in other words each $F_k$ is positive semidefinite and $\sum_{k=1}^r F_k = I$ \cite{holevo1998coding}. In this paper, we make the further assumption that none of the $F_k$'s are zero.  This assumption has always been followed in practice.

There are many examples of entanglement breaking channels, some of which will appear in this paper. Let us note the extreme instance of such a channel, what is called the {\it completely depolarizing channel} $\Phi_{\mathrm{CD}}$, which is defined by $ \Phi_{\mathrm{CD}}(A) = \frac{\tr(A)}{n} I_n$ on $M_n(\mathbb{C})$, and is implemented in the Holevo form with $F_1 = I_n$, $R_1 = \frac1n I_n$. Also, it is given in its rank one form by operators $V_{ij} = \sqrt{n}^{-1} E_{ij}$.

\section{The Holevo Form and Stochastic Matrices}

Let us consider more closely the action of such a map in the form of Eq.~(\ref{holevo}). We note that if $\rho = \Phi(\rho_0)$ is a density matrix in the range of $\Phi$, it belongs to the convex hull of the $R_k$'s and in particular can be written as $\rho = \sum_mc_mR_m$ where the $c_m = \tr(F_m\rho_0)$ are positive real numbers summing to one. Given this decomposition of $\rho$, we get \begin{align*}
\Phi(\rho) &= \sum_k \tr(F_k \rho)R_k \\&= \sum_k \Big(\sum_m c_m \tr(F_kR_m)\Big)R_k,
\end{align*}
and hence $\Phi(\rho) = \sum_ka_kR_k$ where $a_k = \sum_m \tr(F_kR_m)c_m$.

The transformation of the vector $\vec{c} = (c_1 \, c_2 \, \ldots)$ to the vector  $\vec{a} = (a_k)$ can thus be depicted by a matrix multiplication $S \vec{c} = \vec{a}$ in which the $(i,j)$ entry of the matrix $S$ is $\tr(F_iR_j)$; that is,
\[
S = \big( \tr(F_iR_j) \big)_{i,j} .
\]
All entries of $S$ are nonnegative, as $\tr(F_iR_j) = \tr(\sqrt{F_i} R_j \sqrt{F_i}) \geq 0$. Further, the sum of all the entries in any column of this matrix is one since  $\{F_i\}$ forms a POVM and $R_j$ is a density matrix; indeed, for each $j$, we have $\sum_i \tr(F_iR_j) = \tr(\sum_i F_i R_j) = \tr(R_j) = 1.$
Therefore, we have observed that attached to any representation of an entanglement breaking channel in the Holevo form is a (column) stochastic matrix $S = (s_{ij})$
%$$ S = \begin{pmatrix}
%   s_{11} &s_{12} &\hdots &s_{1n} \\   s_{21} &s_{22} &\ddots &s_{2n} \\    \vdots &\ddots &\ddots &\vdots \\   s_{n1} &s_{n2} &\hdots &s_{nn}
%\end{pmatrix}.$$
where $s_{ij}=\tr(F_iR_j)$ for all $i,j$. With the above notation for $\rho$ in the range of $\Phi$ we have $\Phi(\rho) = \sum_k  ( \vec{s}_k \cdot \vec{c}) R_k$, where $\vec{s}_k = (s_{kj} )$ is the $k$th row vector of $S$.

%Now let $V=\{ (c_1,c_2,...,c_n)\in \mathbb{C}^n: \sum_{k=1}^n c_kR_k=0\}$.  It is easy to see that $V$ is an invariant subspace of $S$. Let $W$ be any subspace of $\mathbb{C}^n$ which is complimentary to $V$.  Let $P_{W}$ be the orthogonal projection onto $W$, then the compression $P_W S|_{W}$ is linearly isomorphic to $\Phi$. It follows from this that $rank (\Phi) \le rank(S)$. It also follows from this that any eigenvalue of $\Phi$ must be an eigenvalue of $S$.  The set of all eigenvalues of the $n$ by $n$ stochastic matrices is a proper subset of the closed unit disk and has been characterized by Karpelevich.  The $n$ by $n$ Karpelevich region therefore is also a description of the set of all possible eigenvalues of entanglement breaking channels having a Holevo form with $n$ or less terms.

%\begin{proposition}
%$V$ is an invariant subspace of $S$.
%\end{proposition}

%\begin{proof}
%We define $V$ to be $\{(c_1,c_2, ..., c_n) \in \mathbb{C}^n : \sum_{k=1}^n c_kR_k = 0\}$ and a stochastic matrix S.  We have $S \vec{c} =\sum_{i=j}^n s_{ij} c_j = b_k$.
%\begin{align*}
%\sum_{k=1}^n b_kR_k &= \sum_k(\sum_{j=1} s_{kj}c_j) R_k \\ & = \sum_k \sum_j Tr(F_kR_j)c_j R_k \\ & = \sum_jc_j \Phi(R_j) \\ & = 0
%\end{align*}
%\end{proof}

This stochastic matrix depends on our choice of the Holevo form and therefore is not uniquely determined by the channel.
This leads to a natural question: what properties of the stochastic matrix are invariant under the choice of Holevo form? The following result shows how the structure of any such stochastic matrix representation is closely related to the structure of the channel itself. Note that by the Jordan canonical form of $\Phi$, we mean the (unique) Jordan form of any $n^2\times n^2$ matrix representation of $\Phi$ as an operator on the $n^2$-dimensional space $M_n(\mathbb{C})$.

\begin{theorem}\label{jordanformresult}
Let $\Phi$ be an entanglement breaking channel on $M_n( \mathbb{C})$. Suppose $S$ is a stochastic matrix defined by operators $F_k, R_k$ that define a Holevo form for $\Phi$. Then the Jordan canonical forms of $\Phi$ and $S$ are the same, except possibly on blocks that correspond to zero eigenvalues.
\end{theorem}

\begin{proof}
Suppose we have the Holevo representation $\Phi(\rho) = \sum_{k=1}^r \tr(F_k\rho) R_k$.
We begin by looking at $\Phi$ as a linear operator on $M_n( \mathbb{C})$. Consider a matrix representation $[\Phi]$ of $\Phi$ in a fixed basis for $M_n(\mathbb{C})$ of the matrix units $E_{ij} = \kb{i}{j}$ defined by a fixed orthonormal basis for $\mathbb{C}^n$. For $X\in M_n(\mathbb{C})$ let $\mathrm{vec}(X)$ be the $n^2$-tuple of coordinates for $X$ in this basis viewed as a column matrix. Put
\[
R_k = \sum_{i,j} r_{ij}^{(k)} E_{ij} \quad \mathrm{and} \quad F_k = \sum_{i,j} f_{ij}^{(k)} E_{ij}
\]
for some scalars $r_{ij}^{(k)}, f_{ij}^{(k)}$, so that $\bra{i}F_k\ket{j} =  f_{ij}^{(k)}$, $\bra{i}R_k\ket{j} =  r_{ij}^{(k)}$, and  $\mathrm{vec}(R_k) = (  r_{ij}^{(k)} )_{i,j}$, $\mathrm{vec}(F_k) = (  f_{ij}^{(k)} )_{i,j}$.

Now let $A$ be the $n^2 \times r$ matrix whose $k$th column is $\mathrm{vec}(R_{k})$ for all $1 \leq k \leq r$, and let $B$ be the $r \times n^2$ matrix whose $k$th row is the transposed column matrix $\mathrm{vec}(F_{k}^T)^T$.

We first claim that  $[\Phi] = AB$. Indeed, note that the $s_1 = (i_1,j_1)$, $s_2=(i_2,j_2)$ entry of the $n^2 \times n^2$ matrix $AB$ is, $(AB)_{s_1s_2} = \sum_{k=1}^r r_{i_1j_1}^{(k)} f_{j_2i_2}^{(k)}$, whereas the $s_1,s_2$ entry of $[\Phi]$ is given by,
\begin{eqnarray*}
\bk{\Phi(E_{i_2j_2})}{E_{i_1j_1}} &=& \sum_{k=1}^r \tr (F_k E_{i_2j_2} ) \tr(E_{j_1i_1}R_k) \\
&=& \sum_{k=1}^r  \bra{j_2}F_k\ket{i_2} \bra{i_1}R_k\ket{j_1} \\
&=& (AB)_{s_1s_2}.
\end{eqnarray*}

On the other hand, we claim that the stochastic matrix $S$ defined by the operators $F_k, R_k$ satisfies $S=BA$. To see this, fix a pair $1 \leq k_1,k_2 \leq r$ and observe that
%since $R_k^* = R_k$ we have,
\[
(BA)_{k_1k_2} = \mathrm{vec}(F_{k_1}^T)^T\mathrm{vec}(R_{k_2}) = \sum_{i,j} f_{ij}^{(k_1)} r_{ji}^{(k_2)}= \tr(F_{k_1} R_{k_2} )=(S)_{k_1k_2}.
\]
%whereas we also have,
%\begin{eqnarray*}
%(S)_{k_1k_2} &=& \tr(F_{k_1} R_{k_2} ) \\
%&=& \tr \Big( \sum_{i_1,j_1,i_2,j_2} f_{i_1j_1}^{(k_1)} r_{i_2j_2}^{(k_2)} E_{i_1j_1} E_{i_2j_2} \Big)    \\
%&=& \sum_{i_1,j_2,j} \tr \big(  f_{i_1j}^{(k_1)} r_{jj_2}^{(k_2)} E_{i_1j_2}  \big) \\
%&=&  (BA)_{k_1k_2} .
%\end{eqnarray*}

Thus we have $[\Phi] = AB$ and $S = BA$, and we can now apply the classical Flanders Theorem~\cite{flanders} that relates the Jordan forms of matrix products $AB$ and $BA$ as claimed in the theorem statement.
\end{proof}

The following is an immediate consequence of this result.

\begin{corollary}
Let $\Phi$ be an entanglement breaking channel on $M_n( \mathbb{C})$. Then the non-zero spectrum of $\Phi$ and that of any of its stochastic matrix representations $S$ are the same, including multiplicities.
\end{corollary}

\begin{remark}
More than this, as the long term behaviour of repeated applications of a finite-dimensional linear operator are determined by its Jordan canonical form, it also follows that iterations of $\Phi$ can be modeled by repeated applications of any choice of $S$. We shall return to a special case of this topic in the next section.
\end{remark}

\subsection{Examples}

%\begin{examples}
The extreme case given by the completely depolarizing channel on $M_n(\mathbb{C})$ satisfies $\Phi_{\mathrm{CD}}(\rho) = \frac{\tr(\rho)}{n} I$ for all $\rho$. So in this case we have Holevo operators $F_1 = I$ and $R_1 = \frac1n I$. In this trivial case, with the notation of the theorem, $A$ is the $n^2\times 1$ matrix with $n$ entries of $\frac1n$ corresponding to the diagonal matrix units and $0$'s elsewhere, and $B$ is the $1 \times n^2$ matrix with $n$ entries of $1$ in the same coordinate positions and $0$'s elsewhere. Here $S = BA = (1)$ and $[\Phi]=AB$ is the corresponding rank-$1$ matrix representation of the channel.

Another illustrative example is given by the map-to-diagonal channel $\Lambda$ on $M_n(\mathbb{C})$, which replaces off-diagonal entries of a matrix with zeros, written as $\Lambda(\rho) = \mathrm{diag}(\rho)$. Viewed as a map on operators represented as matrices in a fixed orthonormal basis  $\{ \ket{k} \}^n_{k=1}$, this is a special type of q-c channel (see below) with Holevo form given by $F_k = \kb{k}{k} = R_k$. In this case the factored matrices are related as $B = A^*$, with $A$ as the $n^2 \times n$ matrix whose $n$ columns are the $\mathrm{vec}$ representations of the projections $\kb{k}{k}$, and the stochastic matrix construction yields the $n\times n$ identity matrix $S = BA = I$.

\subsection{Quantum-Classical Channels from Stochastic Matrices}\label{qcsub}

Before continuing, we present a converse in a sense of the connection with stochastic matrices uncovered above. A special class of entanglement breaking channels produce (classical) probability distributions from input quantum states via expectation values from a POVM \cite{horodecki2003entanglement}.

\begin{definition}
A channel is quantum-classical (q-c) if it satisfies Eq.~(\ref{holevo}) and each density operator $R_k = \kb{k}{k}$ is a rank one projection with the set $\{ \ket{k} \}_{k=1}^n$ forming an orthonormal basis for $\mathbb{C}^n$.
\end{definition}

Let $S = (s_{ij})$ be an $n \times n$ (column) stochastic matrix, with each $s_{ij}\geq 0$ and $\sum_{i=1}^n s_{ij} =1$ for all $1\leq j \leq n$. Let $\{ \ket{k} \}_{k=1}^n$ be a fixed orthonormal basis for $\mathbb{C}^n$. For each $k$, let $F_k$ be the operator on $\mathbb{C}^n$ with $n\times n$ diagonal matrix representation in the fixed basis and whose diagonal entries form the $k$th row of $S$. That is,
$$
S = \begin{pmatrix} \textendash\textendash\textendash & \mathrm{diag}(F_1) & \textendash\textendash\textendash \\ \textendash\textendash\textendash & \mathrm{diag}(F_2) & \textendash\textendash\textendash \\ & \vdots \\ \textendash\textendash\textendash & \mathrm{diag}(F_k) & \textendash\textendash\textendash \end{pmatrix},
$$
and explicitly, $F_k = \sum_{j=1}^n s_{kj} \kb{j}{j}$ for each $k$. Observe that every $F_k$ is a positive operator and $\sum_{k=1}^n F_k = I$ by construction.

Now let $R_k = \kb{k}{k}$ for $1\leq k \leq n$ and define a q-c channel on $M_n(\mathbb{C})$ by $\Phi_S (\rho) = \sum_{k=1}^n \tr (F_k \rho) R_k$. Finally, observe that if we apply our stochastic matrix construction above to the entanglement breaking channel $\Phi_S$, we get the matrix $S$ back again as follows: for $1 \leq i,j \leq n$, we have
\[
\tr(F_i R_j) = \sum_{k=1}^n s_{ik} \tr(\kb{k}{k} \kb{j}{j}) = s_{ij}.
\]

We also note this class of channels was considered in \cite{sanz2010quantum} in a context that we will investigate in the next section.

\section{Joint Primitivity for Entanglement Breaking Channels and Stochastic Matrix Representations}

In this section, we make use of the stochastic matrix representations above to derive a matrix theoretic characterization of when entanglement breaking channels are primitive. We begin by reviewing some background.

We note that a quantum channel from $M_n(\mathbb{C}) \to M_n(\mathbb{C})$ maps the compact convex set of density matrices to itself. Therefore Brouwer's theorem guarantees that every such quantum channel must have a density operator fixed point.  This fixed point may or may not be globally attractive; for entanglement breaking channels, we can use the theory of stochastic matrices and the connections developed above to study this question.

 There is a well known sufficient condition for a stochastic matrix to have a unique globally attractive fixed point among the probability vectors.

\begin{definition}
A nonnegative matrix $F$ is said to be primitive if there exists a $m\in \mathbb{N}$ such that $F^m$ has all entries positive.  The smallest such $m\in \mathbb{N}$ that accomplishes this is called the index of primitivity of $F$ and is denoted as $p(F)$.
\end{definition}

The following is one of the most important classical results on primitive matrices \cite{horn2012matrix,meyer2000matrix}.

\begin{theorem}[Perron-Frobenius Theorem]
Let $F$ be an $n \times n$ primitive matrix. Then there is a positive number $\lambda_{max}$ such that all other eigenvalues of $F$ satisfy $|\lambda| < \lambda_{max}$ and the eigenspace associated with $\lambda_{max}$ is one-dimensional. Moreover, there is an eigenvector $v$ of $F$ with eigenvalue $\lambda_{max}$ such that all coordinates of $v$ are positive, and any nonnegative eigenvector of $F$ is a multiple of $v$.
%	\item A Perron vector is the vector $p$defined by $Fp = \lambda_{max}p$, $p>0$, and $||P||_1 = 1$. If $F$ is primitive, then except for positive multiples of p, there are no other nonnegative eigenvectors for A, regardless of the eigenvalue,.
%	\item $\lambda_{max}$ is the only eigenvalue on the spectral circle of F.
%	\item $\lambda_{max} = max_{x \in \mathbb{N}} f(x)$  (Collatz-Wielandt formula)
\end{theorem}

The corollary of this result is that if the associated stochastic matrix (which will satisfy $\lambda_{max} = 1$) of an entanglement breaking channel is primitive then its fixed point is globally attractive.  We note the following equivalent condition for a nonnegative matrix to be primitive.

\begin{proposition} Let $A$ be a nonnegative matrix inside $M_n(\mathbb{R})$.  Then $A$ is primitive if and only if for every nonnegative nonzero $x\in \mathbb{R}^n$, there exists $m>0$ such that $A^m x$ has all of its entries strictly positive.  \end{proposition}

Motivated by this, the concept of a primitive quantum channel was introduced in \cite{sanz2010quantum} as follows (see also \cite{rahaman2019new} for more recent work).

\begin{definition} Let $\Phi$ be a quantum channel, then $\Phi$ is said to be primitive if there exists $m>0$ such that $\Phi^m(\rho)$ is positive definite for all density matrices $\rho$. The smallest $m\in \mathbb{N}$ which accomplishes this is called the index of primitivity of $\Phi$ and is denoted as $q(\Phi)$.
\end{definition}

We present the following result connecting these two notions of primitivity for entanglement breaking channels. The proof relies on the stochastic matrix representations discussed in the previous section. Recall that we have made the standard assumption that none of the $F_k$'s in the POVM in the Holevo form are zero.

\begin{theorem}\label{primitivethm}
Let $\Phi$ be an entanglement breaking channel and $S$ be the stochastic matrix representation associated to the Holevo form $\Phi(X)=\sum_k \tr(F_kX)R_k$.  Then $\Phi$ is a primitive channel if and only if $S$ is a primitive stochastic matrix and $\sum_k R_k$ is positive definite.
\end{theorem}

\begin{proof}
Suppose $S$ is primitive with index of primitivity $m$ and $\sum_k R_k$ is positive definite.  Let $\rho$ be an arbitrary density matrix and let $w$ be the vector with entries $\tr(F_k\rho)$ so $\Phi(\rho)=\sum_k w_k R_k$.  Now let $v=S^m w$ and note that all entries of $v$ must be strictly positive. Since we have the inequalities $\Phi^{m+1}(\rho)=\sum_k v_k R_k \geq (\min_k v_k) \sum_k R_k>0$, it follows that $\Phi$ is a primitive quantum channel with index of primitivity at most $m+1$.

%Then let $v = (v_k)$ be the Perron-Frobenius eigenvector of $S$ with positive coordinates and normalized so that its entries sum to one.  Then for any density matrix $\rho$, using the matrix representation $S$ of $\Phi$ on its range from the previous section we have
%\[ \lim_{m\to \infty}\Phi^m(\rho)=\sum_k v_k R_k.\]

For the converse suppose $\Phi$ is primitive and let $m=q(\Phi)$.  Then for any $j$,  $\Phi^m( R_j)$ is positive definite.  Let $w=S^me_j$, then $\Phi^m(R_j)=\sum_k w_k R_k$.  Since
\[
(\max_k w_k) \sum_k R_k \geq  \sum_k w_k R_k>0,
\]
it follows that $\sum_k R_k$ is positive definite.  Now let $x=Sw=S^{m+1}e_j$.  Then
\[
x_i=\sum_k \tr(F_iR_k)w_k= \tr(F_i(\sum_k w_kR_k)).
\]
As $\sum_{k}w_kR_k=\Phi^m(R_j)$ is positive definite and $F_i$ is (nonzero) positive semidefinite, we have $x_i>0$ and $S^{m+1}e_j$ has all positive entries.  Since $j$ was arbitrary, it follows that $S$ is primitive with index of primitivity less than or equal to $m+1$.
\end{proof}

\begin{remark}
Observe that both conditions in the hypotheses of the theorem are indeed required to describe primitivity of an entanglement breaking channel. If $R = \sum_k R_k$ is not positive definite, then its nullspace is non-zero, and hence the intersection of the nullspaces of the $R_k$ is non-zero (as each $R_k \leq R$), which implies that for any $\rho$, the density operator $\Phi(\rho)$ is not invertible and $\Phi$ cannot be primitive. Moreover, the quantum-classical channels discussed in the previous section, those implemented by a stochastic matrix (which will also be the matrix obtained through the Holevo representation), will not be primitive if the matrix is not primitive, given how the channel's iterative behaviour is so closely tied to that of the matrix for that subclass of channels.
\end{remark}

Let us point out a large class of entanglement breaking channels that satisfy the conditions of the theorem.

\begin{corollary}\label{corprimitive}
Let $\Phi(\rho) = \sum_k \tr(F_k \rho) R_k$ be an entanglement breaking channel such that $\sum_k R_k$ and $F_k$, for all $k$, are positive definite operators. Then $\Phi$ is a primitive channel.
\end{corollary}

\begin{proof}
Each $F_k$ being positive definite (and hence also invertible) together with each $R_k$ being a density matrix, implies $\tr(F_j R_k ) > 0$ for all $j,k$. Thus, the associated stochastic matrix $S$ is primitive, and so by the theorem $\Phi$ is a primitive channel.

Note in this case the primitivity indices are both equal to one; the stochastic matrix $S$ satisfies $p(S)=1$ essentially by definition, and the channel satisfies $q(\Phi)=1$ as $\Phi(\rho)$ will be positive definite for every density matrix $\rho$ from the given properties of the $F_k$ and $\sum_k R_k$.
\end{proof}

Based on the proof of the theorem above, we can also give the following statement on the relationship between the primitivity indices of a channel and its stochastic matrices.

\begin{corollary}\label{primindex}
Let $\Phi$ be a primitive entanglement breaking channel and let $S$ be one of its (primitive) stochastic matrix representations.  Then,
\[
\vert q(\Phi)-p(S) \vert\le 1.
\]
\end{corollary}

\begin{proof}
If $S$ is primitive, and $\sum_k R_k$ is positive definite, then the proof of Theorem~\ref{primitivethm} shows that $q(\Phi) \leq p(S)+1$. On the other hand, if $\Phi$ is primitive, then that
proof also shows that $p(S) \leq q(\Phi)+1$.
\end{proof}

\begin{remark}
We note that equality of these two primitivity indices is satisfied for the channels of Corollary~\ref{corprimitive}, where $q(\Phi) = p(S) =1$. The same is true for the primitive channels from the special class of quantum-classical channels considered in the previous section, where primitivity of the channel is easily seen to be  equivalent to primitivity of the defining stochastic matrix via Theorem~\ref{primitivethm}. This was first observed in  \cite{sanz2010quantum}, where the two notions were also connected with the size of operator spaces spanned by products of the channel's Kraus operators.
\end{remark}

The two primitivity indices can be different, however, as the following pair of examples show.

\subsection{Examples}

%\begin{examples}

Consider the `single-qubit' entanglement breaking channel $\Phi : M_2(\mathbb{C}) \rightarrow M_2(\mathbb{C})$ with Holevo form defined by $R_1 = \kb{0}{0}$, $R_2 = \kb{1}{1}$, and $F_1 = \kb{+}{+}$, $F_2 = \kb{-}{-}$ where $\ket{\pm} =\frac{1}{\sqrt{2}}(\ket{0} \pm \ket{1})$. Then observe that
\[
S = \big(\tr(F_i R_j) \big)_{i,j} = \left( \begin{matrix}
\frac12 & \frac12 \\
\frac12 & \frac12
\end{matrix} \right),
\]
and so $S$ is primitive with $p(S)=1$. We also have $R_1+R_2=I$ is positive definite, and hence by the theorem above we know $\Phi$ is primitive. However, note that $\Phi(\kb{-}{-}) = \kb{1}{1}$ is not positive definite, and so $q(\Phi) \gneq 1$. In fact, we know $q(\Phi) =2$ from the above corollary as $q(\Phi) \leq p(S) + 1 = 2$. We can observe this directly by computing that for all single-qubit density matrices $\rho$, we have $\Phi^2(\rho) = \frac12 I$, and hence $\Phi^2$ is the completely depolarizing channel which is of course primitive.

On the other hand, we can show the other primitivity index inequality is sharp by using the following Holevo form for the single-qubit completely depolarizing channel $\Phi= \Phi_{\mathrm{CD}}$. Let $F_1 = \frac12 \kb{0}{0}$, $F_2 = \frac12 \kb{1}{1}$, $F_3 = \frac12 I$, and $R_1 = R_2 = \kb{0}{0}$, $R_3 = \kb{1}{1}$. Thus, $q(\Phi) = 1$. However, the stochastic matrix associated with this decomposition is
\[
S = \big(\tr(F_i R_j) \big)_{i,j} = \left( \begin{matrix}
\frac12 & \frac12 & 0 \\
0 & 0 & \frac12 \\ \frac12 & \frac12 & \frac12
\end{matrix} \right),
\]
and so $p(S) \gneq 1$. The inequality of the corollary tells us $p(S) = 2$ as $p(S) \leq q(\Phi) +1 = 2$, and this can be observed directly by noting that $S^2 > 0$.

%\end{examples}

\subsection{Holevo Rank of an Entanglement Breaking Channel}

It follows that our result relating the primitivity indices of an entanglement breaking channel and its stochastic matrix representations is optimal for the full class of such channels, in the sense that it identifies bounds between the indices which can be saturated. These indices are related to the following notion for entanglement breaking channels, which is natural to consider, but does not appear to have been formally investigated previously.

\begin{definition}
Given an entanglement breaking channel $\Phi$, we define the Holevo rank of the channel to be the minimal number $r$ of pairs $\{F_k, R_k\}_{k=1}^r$ that make up a Holevo form for the channel, $\Phi(\rho) = \sum_{k = 1}^r \tr(F_k \rho) R_k$.
\end{definition}

As a class of channels, entanglement breaking channels can be characterized by properties of their Choi matrix; namely, a channel is entanglement breaking if and only if its Choi matrix is not entangled, or separable. Analogous to this property, we can also frame the Holevo rank in terms of properties of the Choi matrix. Indeed, observe that if $\Phi$ has a Holevo form determined by $\{F_k, R_k\}_{k=1}^r$, then the Choi matrix satisfies:
\[
J(\Phi) = \sum_{k=1}^r \tau(F_k) \otimes R_k,
\]
where $\tau$ is the transpose map. Thus, the Holevo rank is also the minimal size of such a sum decomposition of the Choi matrix, with the $F_k$ forming a POVM and each $R_k$ a density matrix.

We can use the stochastic matrix representations and our results above to bound the primitivity index of an entanglement breaking channel in terms of the Holevo rank as follows.

\begin{corollary}\label{primindexupperbound}
Let $\Phi$ be a primitive entanglement breaking channel on $M_n(\mathbb{C})$ with Holevo rank $r$. Then we have,
\[
q(\Phi) \leq r^2 -2r + 3.
\]
\end{corollary}

\begin{proof}
First recall the classical Wielandt inequality \cite{wielandt1950unzerlegbare,meyer2000matrix,horn2012matrix}, which for a $r \times r$ primitive matrix $S$ gives $p(S) \leq r^2 -2r +2$. The result thus immediately follows from Corollary~\ref{primindex}.
\end{proof}

\begin{remark}
One could compare this upper bound to the quantum version of Wielandt's inequality from \cite{sanz2010quantum}, which was established for general (not necessarily entanglement breaking) channels as: $q(\Phi) \leq (n^2 - d + 1) n^2$, where $d$ is the number of Kraus operators required to implement the channel. Additionally, further investigation is warranted on the relationship between these quantities and the recently studied notion of entanglement breaking rank \cite{pandey2020entanglement}.
\end{remark}

\section{Conclusion \& Outlook}

We showed that every entanglement breaking channel naturally determines a stochastic matrix from each of its Holevo form representations. While the matrix representation is not unique, we proved that any such matrix has non-zero spectrum that is the same as the channel viewed as a linear map. It would be interesting to determine whether the channel-stochastic matrix representation is continuous, in the sense that two such channels are close if and only if there are two matrix representations for the channels that are close, appropriately measured.

We used the stochastic matrix representations to investigate primitivity for entanglement breaking channels, and we discovered a computable and direct relationship between primitivity of the channel on the one hand, and primitivity of any of its stochastic matrix representations together with a condition on the density matrices from the Holevo form on the other. This led to the identification of a number of subclasses of examples and an explicit relationship between the primitivity indices of a  channel and its matrix representations. It might be worth considering potential further applications of this channel and matrix primitivity relationship in other quantum information settings that involve entanglement breaking channels.

We also introduced the notion of Holevo rank for entanglement breaking channels and used our results to bound a channel's primitivity index in terms of this rank. We feel further investigation of this rank is warranted, in particular as to how it might be related to other quantities associated with such channels, or the Choi matrix for the channels.

More generally, it could be interesting to consider other results from the theory of stochastic matrices, and see if they can also be exported to produce new results in the theory of entanglement breaking channels via the matrix representations we have identified here. We plan to undertake these investigations elsewhere.

\vspace{0.1in}

{\noindent}{\it Acknowledgements.} We are grateful to the referee for helpful comments and identification of some corrections. J.A. was partly supported by a Mitacs Accelerate internship at the African Institute for Mathematical Sciences (AIMS). D.W.K. was partly supported by NSERC. R.P. was partly supported by NSERC.

\bibliographystyle{plain}
\bibliography{AKLPRbib}

%\bibliography{privacy-mult}
%\bibliographystyle{amsplain}

\end{document}